\documentclass[%showpacs,
showkeys,12pt,
preprint,preprintnumbers,nofootinbib,
groupedaddress,superscriptaddress,amsmath,amssymb]{revtex4}
\usepackage{graphicx}% Include figure files
\usepackage{dcolumn}% Align table columns on decimal point
\usepackage{bm}% bold math
\usepackage{amssymb}
\usepackage{amsmath}
\usepackage{epsfig}    
\usepackage{color}
\usepackage{slashed}
\usepackage{hhline}
\def\be{\begin{equation}}
\def\ee{\end{equation}}
\newcommand{\bea}{\begin{eqnarray}}
\newcommand{\eea}{\end{eqnarray}}
\newcommand{\nn}{\nonumber}

\numberwithin{equation}{section}

\newcommand{\ET}{\mbox{$\not \hspace{-0.10cm} E_T$ }}

\begin{document}

%%%%%%%%%
%\title{Two loop neutrino model linking to leptoquark and diquark \\ with gauged hidden $U(1)$ symmetry}
\title{A Colored KNT Neutrino Model }
%\preprint{KIAS-P14078}
%

\author{Takaaki Nomura}
\email{nomura@kias.re.kr}
\affiliation{School of Physics, KIAS, Seoul 130-722, Korea}

\author{Hiroshi Okada}
\email{macokada3hiroshi@cts.nthu.edu.tw}
\affiliation{Physics Division, National Center for Theoretical Sciences, Hsinchu, Taiwan 300}

\author{Nobuchika Okada}
\email{okadan@ua.edu}
\affiliation{Department of Physics and Astronomy, University of Alabama, Tuscaloosa, AL35487, USA}

\date{\today}

\begin{abstract}
We propose a radiative seesaw model at the three-loop level, in which quarks, leptons, leptoquark bosons, and a Majorana fermion of dark matter candidate are involved in the neutrino loop.
Analyzing neutrino oscillation data including all possible constraints such as flavor changing neutral currents, lepton flavor violations, upper/lower bound on the mass of leptoquark from the collider physics, and the measured relic density of the dark matter,
we show the allowed region to satisfy all the data/constraints. 
\end{abstract}
\maketitle
\newpage

\section{Introduction}
Since it is experimentally proved that neutrino masses are very tiny compared to the other three fermion sectors in the standard model (SM), one often considers new mechanisms to induce such a tiny neutrino masses naturally.
One of the promising scenarios is to radiatively generate neutrino masses by forbidding the tree-level masses that is sometimes called
{\it radiative seesaw models}, and there are a lot of papers along this idea. For example, one loop induced models are found in Ref.~\cite{onelps}, two-loop ones are found in Ref.~\cite{twolps}, three-loop ones are found in Ref.~\cite{threelps}, and
see Ref.~\cite{Fourlps} for four-loop ones.
Especially, if known particles such as quarks and leptons are simultaneously running inside the neutrino loop,
we could interpret the known SM fermions play an important role in providing the tiny neutrino masses and
more variety of phenomenologies such as flavor changing neutral currents (FCNCs), lepton flavor violations (LFVs), muon anomalous magnetic moment, electric dipole  moment, can potentially be taken into account as well as the neutrino oscillation data.
To achieve such kinds of models, leptoquark (LQ) bosons, which have $SU(3)$ color degrees of freedom in the SM gauge symmetry, are needed to connect each others. This line of ideas is found in Ref.~\cite{Kohda:2012sr, Dasgupta:2013cwa, Nomura:2016ask}.
%%% %%%
In another aspect of the {\it radiative seesaw models}, a dark matter (DM) candidate is often involved in the neutrino loop.
One of the reasons is that DM should be electrically neutral and tends to be weakly interacting particle. Therefore, the nature of DM is similar to the active neutrinos (if DM is especially fermion), and it could be natural to consider that these particles are correlated with each other.
Moreover, the mass scale of DM is not confirmed yet although many experiments are running to search for the DM candidate.
In this sense, its mass can be treated as a free parameter to fit the neutrino oscillation data as well as the other phenomenologies.

In this paper, we propose a radiative seesaw model  at the three-loop level that possesses all the contents discussed above.
Here, all the (down-type) quarks, leptons, LQ, and DM, are mediated inside the neutrino loop.~\footnote{
Its framework is however already discussed in Ref.~\cite{Chen:2014ska} as one of the possibilities of such an radiative neutrino model.}
Then we analyze neutrino oscillation data including all the possible constraints coming from quarks, leptons, LQ, and DM,
and show the allowed region to satisfy all the data. 

%This is an extension of a colored Zee-Babu model~\cite{oka-nomu} based on an extended colored Zee-Babu model~\cite{oka-nomu} to include a Majorana fermionic dark matter (DM) candidate, and neutrino masses are induced at the three-loop model with Krauss-Nasri-Trodden (KNT) type of model~\cite{knt}.

This paper is organized as follows.
In Sec.~II, we show our model,  including neutrino mass matrix.
In Sec.~III, we discuss phenomenology of the model such as flavor violation, dark matter and collider physics, and show numerical results to satisfy all the data.
Sec.~IV is devoted for conclusions and discussions.
%We conclude and discuss in Sec.~IV.

%\newpage

\section{Model}
In this section, we introduce our model including formula of active neutrino mass matrix.
%%%%%%%%%%%%%%%%%%%%%%%%%%%%%%%%%%%%%
 \begin{widetext}
\begin{center} 
\begin{table}%[tbc]
%\begin{tiny}
\begin{tabular}{|c||c|c|c||c|c||c|}\hline\hline  
&\multicolumn{3}{c||}{Quarks} & \multicolumn{2}{c||}{Leptons}& \multicolumn{1}{c|}{Dark Matter} \\\hline
& ~$Q_{L_i}$~ & ~$u_{R_i}$~ & ~$d_{R_i}$ ~ 
& ~$L_{L_i}$~ & ~$e_{R_i}$ ~ & ~$N_{R_i}$ 
\\\hline 
%%%
$SU(3)_C$ & $\bm{3}$  & $\bm{3}$  & $\bm{3}$  & $\bm{1}$& $\bm{1}$& $\bm{1}$   \\\hline 
%%%
$SU(2)_L$ & $\bm{2}$  & $\bm{1}$  & $\bm{1}$  & $\bm{2}$& $\bm{1}$& $\bm{1}$   \\\hline 
 %%%
$U(1)_Y$ & $\frac16$ & $\frac23$  & $-\frac{1}{3}$  & $-\frac12$  & $-1$  & $0$\\\hline
 %%%
% $U(1)_H$ & $0$  & $0$ & $0$  & $x$ & $x$ & $2x$   \\\hline
%%%%
$Z_2$ & $+$ & $+$  & $+$ & $+$ & $+$ & $-$\\\hline
%%%
\end{tabular}
\caption{Field contents of fermions
and their charge assignments under $SU(3)_C\times SU(2)_L\times U(1)_Y\times Z_2$, where the lower index $i(=1-3)$ represents flavors. }
\label{tab:1}
% \end{tiny}
\end{table}
\end{center}
\end{widetext}
\begin{table}%[thbp]
\centering {\fontsize{10}{12}
\begin{tabular}{|c||c|c|c|}\hline\hline
&~ $\Phi$  ~&~ $S^{a}_{\rm LQ_1}$  ~&~ $S^{a}_{\rm LQ_2}$ \\\hline
%%%
$SU(3)_C$ & $\bm{1}$  & $\bm{3}$ & $\bm{3}$ \\\hline 
$SU(2)_L$ & $\bm{2}$   & $\bm{1}$  & $\bm{1}$  \\\hline 
$U(1)_Y$ & $\frac12$  & $-\frac{1}{3}$ & $-\frac{1}{3}$   \\\hline
% $U(1)_H$ & $0$ & $-x$  & $-2x$  & $x$ & $4x$   \\\hline
$Z_2$ & $+$   & $+$  & $-$ \\\hline
\end{tabular}%
} 
\caption{Field contents of bosons
and their charge assignments under  $SU(3)_C\times SU(2)_L\times U(1)_Y\times Z_2$. }
\label{tab:2}
\end{table}
\subsection{ Model setup}
We show all the field contents and their charge assignments in Table~\ref{tab:1} for the fermion sector and  Table~\ref{tab:2} for the boson sector. 
Under this framework, the relevant part of the renormalizable Lagrangian and the Higgs potential are given by
{
\begin{align}
\label{eq:Yukawa}
-{\cal L}^{}&=
(y_\ell)_{ij}\bar L_L \Phi e_{R_j} + (y_L)_{ij}\bar L^c_{L_i}(i\sigma_2) Q_{L_j}^\alpha S_{LQ_1}^{*\alpha}
+(y_S)_{ij}\bar d^{c\alpha}_{R_i} N_{R_j} S_{DQ_2}^{*\alpha} + M_{N_i} \bar N_{R_i}^c N_{R_i}+{\rm h.c.} ,\\
%%%
{\cal V}&=m_\Phi^2 \Phi^\dag \Phi + m_{S_{LQ_1}}^2 S_{LQ_1}^{*\alpha} S_{LQ_1}^{\alpha}
+ m_{S_{LQ_2}}^2 S_{LQ_2}^{*\alpha} S_{LQ_2}^{\alpha} \nn\\
& +\lambda_0[( S_{LQ_1}^{*\alpha} S_{LQ_2}^\alpha)( S_{LQ_1}^{*\beta} S_{LQ_2}^\beta)+{\rm c.c.}]
+\lambda_0'[( S_{LQ_1}^{*\alpha} S_{LQ_2}^\beta)( S_{LQ_1}^{*\beta} S_{LQ_2}^\alpha)+{\rm c.c.}]\nn\\
%%%
&+\lambda_0''[( S_{LQ_1}^{*\alpha} S_{LQ_2}^\beta)( S_{LQ_1}^{*\beta} S_{LQ_2}^\alpha)+{\rm c.c.}]
+\lambda_{\Phi} |\Phi^\dag \Phi|^2
+\lambda_{S_{LQ_1}} |S_{LQ_1}^{*\alpha} S_{LQ_1}^\alpha|^2
+\lambda_{S_{LQ_2}} |S_{LQ_2}^{*\alpha} S_{LQ_2}^\alpha|^2 \nn\\
&+\lambda_{\Phi S_{LQ_1}} (\Phi^\dag \Phi)( S_{LQ_1}^{*\alpha} S_{LQ_1}^\alpha)
+\lambda_{\Phi S_{LQ_2}} (\Phi^\dag \Phi)( S_{LQ_2}^{*\alpha} S_{LQ_2}^\alpha)
+\lambda_{S_{LQ_1} S_{LQ_2}} (S_{LQ_1}^{*\alpha} S_{LQ_1}^\alpha)( S_{LQ_2}^{*\beta}S_{LQ_2}^\beta)\nn\\
%%%
&
+\lambda'_{S_{LQ_1} S_{LQ_2}} (S_{LQ_1}^{*\alpha} S_{LQ_1}^\beta)( S_{LQ_2}^{*\alpha}S_{LQ_2}^\beta)
+\lambda''_{S_{LQ_1} S_{LQ_2}} (S_{LQ_1}^{*\alpha} S_{LQ_1}^\beta)( S_{LQ_2}^{*\alpha}S_{LQ_2}^\beta),
\end{align}
where $\sigma_2$ is the second component of the Pauli matrices.
Each of $\lambda'_{S_{LQ_1} S_{LQ_2}}$ and $\lambda''_{S_{LQ_1} S_{LQ_2}} $
comes from the contract of $(3\times3)(\bar3\times\bar3)\to(\bar 3)\times(3)\to 1$, and $(\bar 6)\times(6)\to 1$, where we used $3 \times 3 = \bar 3 + 6$ for $SU(3)$ representations. 
%%%
Also $\lambda_{S_{LQ_1} S_{LQ_2}} $
comes from the contract of $(\bar3\times3)(\bar3\times3)\to1$, and $( 8)\times(8)\to 1$, where we used $\bar3 \times 3 = 1 + 8$ for $SU(3)$ representations.
%For simplicity, we set $\lambda_0\equiv \lambda_{S_{LQ_1} S_{LQ_2}}\simeq\lambda'_{S_{LQ_1} S_{LQ_2}}\simeq\lambda''_{S_{LQ_1} S_{LQ_2}} $
Also $\lambda_0'$ and $\lambda_0''$ comes from the same contract as $\lambda'_{S_{LQ_1} S_{LQ_2}}$ and $\lambda''_{S_{LQ_1} S_{LQ_2}} $. But for simplicity  we set $\lambda_0\equiv \lambda_0'\simeq\lambda_0''$ hereafter. Therefore there exists 15 color factor 
}

\subsection{Active neutrino mass matrix}
The neutrino mass matrix is induced at the three-loop level, and its formula is given by 
\begin{align}
&{\cal M}_{\nu_{ab}}\approx\frac{{15}\lambda_0}{(4\pi)^6 m_{LQ_1}^2} 
(y_L)_{ai} m_{d_i} (y_S^*)_{ij} M_{N_j}(y_S^\dag)_{jk}m_{d_k} (y_L^T)_{kb}
 F_3(r_{N_j}, m_{LQ_2}) ,\\
&F_3(r)=\int [dx]  \int [dx' ]  \int [dx'']  
\frac{\delta(1-x-y-z)\delta(1-x'-y'-z')\delta(1-x''-y''-z'')}{x r_{N_j} - y \Delta_1[r_{d_i},  r_{SL_2}]-z\Delta_2[r_{d_k},r_{SL_2}]},\\&
{
\Delta_1[r_{d_i},r_{SL_2}]=
\frac{x' r_{d_i}+y' +z' r_{SL_2}}{z'^2-z'},\quad
\Delta_2[r_{d_k} , r_{SL_2}]
=\frac{x'' r_{d_k}+y'' + z'' r_{SL_2}}{z''^2-z''},
}
\end{align}
where the factor ${15}$ in the neutrino mass matrix comes from total color-degrees of freedom, $r_f\equiv(m_f/m_{LQ_1})^2$, 
$[dx]\equiv dxdydz$, and one can assume to be $r_{d_i(k)} \simeq0$. 
%%%
Notice here that $F_3(r)$ is derived by directly computing the Feynman integrations, although this form looks different from the standard form found in Ref.~\cite{Ahriche:2013zwa}.  
It is convenient to perform the full analysis including the neutrino oscillation data,
%%%
and its data is given by diagonalizing ${\cal M}_{\nu_{ab}}$ as follows:
\begin{align}
{\cal M}^{diag}_\nu=V_{MNS}^T {\cal M}_{\nu} V_{MNS},
\end{align}
where $V_{MNS}$ is the Maki-Nakagawa-Sakata mixing matrix.
Furthermore, we adopt a method of Casas-Ibarra parametrization~\cite{Casas:2001sr} to carry out our numerical analysis with such a complicated neutrino mass matrix structure. In our case, the parametrization can generally be found as
\begin{align}
y_L&=V_{MNS}^*\sqrt{{\cal M}^{diag}_\nu} {\cal O} A^{-1/2} (y_S^*)^{-1} m_d^{-1},\label{eq:ci-1}\\
{\rm or}\nn\\
y_S&=\left[m_d^{-1}  (y_L)^{-1} V_{MNS}^*\sqrt{{\cal M}^{diag}_\nu} {\cal O} A^{-1/2}\right]^* ,\label{eq:ci-2}
\end{align}
where
\begin{align}
A&\equiv \frac{{15}\lambda_0}{(4\pi)^6 m_{LQ_1}^2} M_N F_3(r_{N}, m_{LQ_2}) ,\quad 
{\cal O}\equiv
 \left[\begin{array}{ccc} 
1 &0 & 0 \\
0 & c_\alpha & s_\alpha \\
0 & -s_\alpha & c_\alpha \\
  \end{array}
\right].
\end{align}
Here, ${\cal O}$ is a complex orthogonal matrix; ${\cal O}^T {\cal O}=1$.
Depending on experimental constraints, one can select more convenient one.
In our case we select the case of Eq.~(\ref{eq:ci-2}),
because $y_L$ has to be imposed  a lot of experimental constraints  than $y_S$.
Therefore, $y_L$ is taken as an input parameter in our numerical analysis.
%%%
For the neutrino oscillation data, we have used the best fit values with the global analysis in Ref.~\cite{Forero:2014bxa};
\begin{eqnarray}
&&  s_{12}^2 = 0.323, \; 
 s_{23}^2 = 0.567, \;
 s_{13}^2 = 0.0234,   \;
\delta_{CP} =1.34 \pi,
\\
&& 
%  m_{\nu_2} ({\rm eV}) = 0.0087,  \; 
  \ |m_{\nu_3}^2- m_{\nu_2}^2| =2.48 \times10^{-3} \ {\rm eV}^2,  \; 
 % m_{\nu_3} ({\rm eV}) = 0.0502 .
  \ m_{\nu_2}^2- m_{\nu_1}^2 =7.60 \times10^{-5} \ {\rm eV}^2, \nn
  \label{eq:neut-exp}
  \end{eqnarray}
where we assume one of three neutrino masses is zero with normal ordering, for simplicity, in the numerical analysis below.
%Here we prepare the specific textures on $y_L$ and $y_S$ to evade some constraints as we will see in the next subsection.

\section{Phenomenology of the model}

In this section, we discuss phenomenology of the model which includes lepton flavor violations, dark matter physics and collider physics.

\subsection{Flavor Changing Neutral Currents and Lepton Flavor Violations}
\if0
Before discussing the Flavor Changing Neutral Currents (FCNCs) and the lepton flavor violations (LFVs),
we specify the textures of $y_L$ and $y_S$ as simple as possible to evade some  FCNC processes:
\begin{align}
y_L= V_{MNS} \times y_L' \equiv V_{MNS}\times \left[\begin{array}{ccc} (y_L)_{11} &0 & 0 \\
 0 & (y_L)_{22} & 0 \\a
0 &0 & (y_L)_{33} \\
  \end{array}
\right],\quad
%%%
y_S=\left[\begin{array}{ccc} (y_S)_{11} &0 & 0 \\
0 & (y_S)_{22} & 0 \\
0 & 0 & (y_S)_{33} \\
  \end{array}
\right].
\end{align}
\fi
%%%
Here we discuss the Flavor Changing Neutral Currents (FCNCs) and the lepton flavor violations (LFVs),
where all the constraints related to $y_L$  are the same as the original colored Zee-Babu model~\cite{Kohda:2012sr, Nomura:2016ask}. 
Thus we just provide the most stringent constraint on $y_L$, which comes from the process of $\mu\to e\gamma$ and its branching ratio is given by
\begin{align}
BR(\mu\to e\gamma)\approx\frac{3\alpha_{em}}{256\pi G_F^2 m_{LQ_1}^4}|(y_Ly_L^\dag)_{21}|^2,
\end{align}
where $\alpha_{em}$ is the fine-structure constant, and $G_F$ is the Fermi constant. 
Current experimental bound is given by~\cite{TheMEG:2016wtm}
\begin{align}
BR(\mu\to e\gamma)_{exp}\lesssim 4.2\times10^{-13}.
\end{align}
%\footnote{Muon anomalous magnetic moment }
%%%
On the other hand, $y_S$ gives  nonzero contributions to $b\to s\gamma$, and $K^0-\bar K^0$ and $B_d^0-\bar B_d^0$ mixings through the one-loop box diagrams.
The (partial) decay rate of $b\to s\gamma$ through the box diagram is given by
\begin{align}
&\Gamma(b\to s\gamma)\approx \frac{\alpha_{em} m_b^5}
{442368\pi^4}\nn\\
&
\left|
\frac{
(y_S)_{3a}(y_S^\dag)_{a2}\left[
m_{LQ_2}^6 - 6m_{LQ_2}^4 M_{N_a}^2 +3m_{LQ_2}^2 M_{N_a}^4+2M_{N_a}^6+12 6m_{LQ_2}^2 M_{N_a}^4\ln\left(\frac{m_{LQ_2}}{M_{N_a}}\right)  \right] }
{(m_{LQ_2}^2-M_{N_a}^2)^4}
\right|^2,
\end{align}
 then the branching ratio is given by 
 \begin{align}
 BR(b\to s\gamma)\approx
&\frac{\Gamma(b\to s\gamma)}{\Gamma_{tot.}} \lesssim 3.29\times 10^{-4}.
\end{align}
 where  $\Gamma_{tot.}\approx 4.02\times10^{-13}$ GeV is the total decay width of bottom quark, and the right side value is the experimental upper bound~\cite{Lees:2012wg}.

The forms of $K^0-\bar K^0$ and $B_d^0-\bar B_d^0$ mixings are, respectively, given by
\begin{align}
&\Delta m_K\approx
\frac{2m_L f_K}{3(4\pi)^2}|(y_S)_{11}|^2|(y_S)_{22}|^2 F_{box}(M_{N_1},M_{N_2},M_{LQ_2})\lesssim 3.48\times10^{-15}[{\rm GeV}],\label{eq:kk}\\
&\Delta m_B\approx
\frac{2m_B f_B}{3(4\pi)^2}|(y_S)_{11}|^2|(y_S)_{33}|^2 F_{box}(M_{N_2},M_{N_3},M_{LQ_2})\lesssim 3.36\times10^{-13}[{\rm GeV}],\label{eq:bb}\\
&F_{box}(m_1,m_2,m_3)
=\int \frac{\delta(1-a-b-c-d)dadbdcdd}{[a m_1^2+b m_2^2+(c+d) m_3^2]^2},
\end{align}
where each of the last inequalities of Eqs.(\ref{eq:kk}, \ref{eq:bb}) represents the upper bound on the experimental values, and $f_K\approx0.156$ GeV, $f_B\approx0.191$ GeV, $m_K\approx0.498$ GeV, and $m_B\approx5.280$ GeV.
\footnote{Since we assume that one of the neutrino masses be zero with normal ordering that leads to the 1st column of $y_S$ is almost zero; $(y_S)_{11}\approx 0$, these constraints can easily be evaded. }

\subsection{Dark Matter}
Here we identify $N_2$ as a DM candidate, and define its mass to be $M_{N_2}\equiv M_X$.~\footnote{In the numerical analysis, we obtain that the 1st column of $y_S$ is almost zero that leads to over relic density. Thus, $N_1$  is not a good DM candidate. }
The DM dominantly annihilate into down type quarks, $N_2 N_2 \to d_i \bar d_j$, via $S_{LQ_2}$ exchange by interaction with coupling $y_S$.
The relic density is approximately given by
\begin{align}
\Omega h^2\approx \frac{4.28\times10^{9} x_f^2 }
{\sqrt{g^*} M_P [a_{eff}(-3+4 x_f)+12b_{eff}]},
\end{align}
where $g^*\approx100$, $M_P\approx 1.22\times 10^{19}$, $x_f\approx25$, and
its effective s-wave and p-wave in the limit of massless final state of down type quarks are, respectively, given by
\begin{align} 
a_{eff}& \simeq 0, \\
%\frac{|(y_S^\dag y_S)_{22}|^2}{64 \pi}\frac{ M_{X}^2 }{(M_{X}^2 + m_{LQ_2}^2)^2},\\
b_{eff}& 
{
\simeq \frac{|(y_S^\dag y_S)_{22}|^2}{64 \pi}  \frac{M_X^2 (m_{LQ_2}^4 + M_X^4)}{(m_{LQ_2}^2+m_X^2)^4}.
}
%\frac{|(y_S^\dag y_S)_{22}|^2}{192 \pi} \frac{ M_{X}^4 (m_{LQ_2}^4-M_{X}^4) -3 M_{X}^6m_{LQ_2}^2}{M_{X}^2 (M_{X}^2 + m_{LQ_2}^2)^4 }.
\end{align}
%where we took the limit $m_{d_i} \to 0$ since we assume DM is much heavier than down type quarks.
Note that the s-wave contribution is suppressed since it is proportional to square of down type quark mass.
In our numerical analysis below, we use the current experimental range approximately as $0.11\le \Omega h^2\le 0.13$~\cite{Ade:2013zuv}.

\subsection{Numerical analysis \label{sec:numerical}}
\begin{figure}[t]
\begin{center}
\includegraphics[width=70mm]{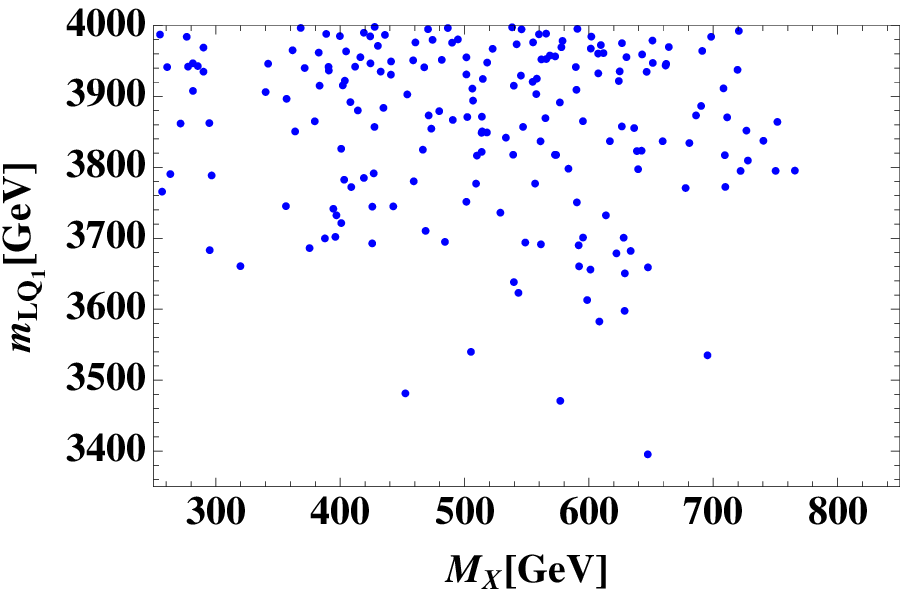}
\includegraphics[width=70mm]{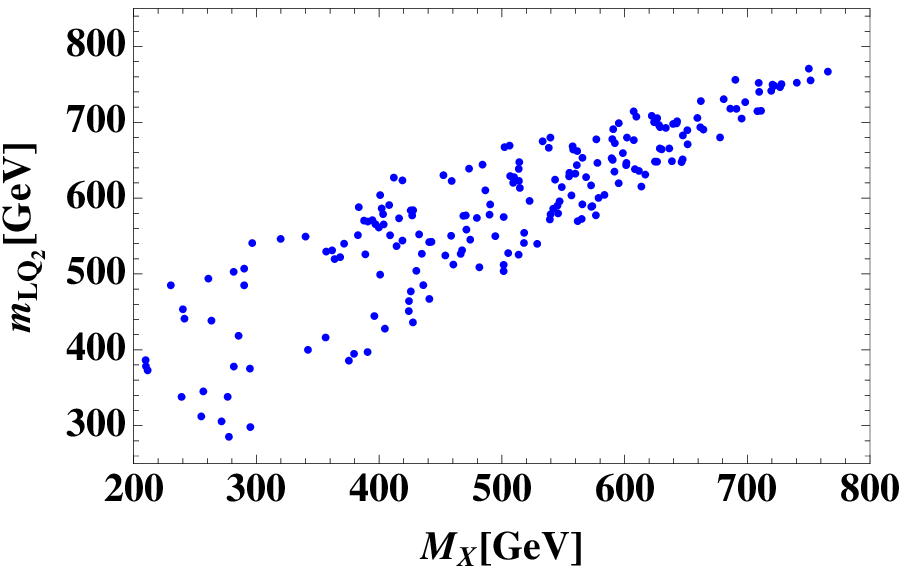}
\caption{Allowed regions to satisfy all the constraints such as neutrino oscillation data, LFVs, FCNC, and the measured relic density.
The left figure represents the allowed region in terms of the mass of DM and $S_{LQ_1}$,
and the right figure represents the allowed region in terms of the mass of DM and $S_{LQ_2}$. }
\label{fig:x-ql1}
\end{center}
\end{figure}

Here, we search for the allowed region to satisfy all the constraints such as LFVs, FCNCs, and the relic density of DM that have already been  discussed above. First of all we fix the range of  input parameters as follows:
\begin{align}
& M_{X} \in [200\,, 800\,] \; \text{GeV},\quad \{M_{N_3}, M_{N_1} \} \in [M_X\,, 4000\,]\; \text{GeV}, \nn \\
& m_{LQ_1} \in [2500\,, 4000\,]\; \text{GeV}, \quad m_{LQ_1} \in [M_X\,, 1000\,]\; \text{GeV}, \nn\\
%\; \text{GeV},\quad  \mu \in [500\,, 3000]\ \text{GeV}, \nn\\
&  (y_L)_{11} \in [0.02\,, 0.05]\, , \quad  (y_L)_{12} \in [0.013\,, 0.02]\, , \quad  (y_L)_{13} \in [0.003 \,,0.005],\nn\\
&  (y_L)_{21} \in [0.01\,, 0.05]\, , \quad  (y_L)_{22} \in [0.1\,, 0.2]\, , \quad  (y_L)_{23} \in [0.0019 \,,0.005],\nn\\
&  (y_L)_{31} \in [0.017\,, 0.020]\, , \quad  (y_L)_{32} \in [0.014\,, 0.020]\, , \quad  (y_L)_{33} \in [0.29 \,,0.50],\nn\\
&  \alpha \in [(-1-i)/1000\,,(1+i)/1000],
	\label{eq:range_scanning}
\end{align} 
where  $\lambda_0=4\pi$ and LFVs require rather small $y_L$.~\footnote{Since lager values of $y_L$ do not results in an allowed region,
we have chosen such a specific region.}
We also find that the mass of $S_{LQ_2}$ is preferred to be lighter than 1 TeV while that of $S_{LQ_1}$ is required to be heavy as several TeV.
The 5 million random parameter sets are applied for numerical calculation and the results are shown in Fig.~\ref{fig:x-ql1}, where 202 points satisfy all the constraints.
The left plot of Fig.~\ref{fig:x-ql1}, represents the allowed region in terms of the mass of DM and $S_{LQ_1}$.
One finds that smaller mass of $S_{LQ_1}$ is not allowed. This mainly comes from the constraint of LFVs such as $\ell_i\to \ell_j\gamma$.
%%%
On the other hand, the right plot of Fig.~\ref{fig:x-ql1} represents the allowed region in terms of the mass of DM and $S_{LQ_2}$
that gives the upper bound on the mass of $S_{LQ_2}$, $m_{LQ_2}\lesssim 800$ GeV.
%Moreover, the larger mass of DM, the narrower the allowed region becomes. 
This constraint mainly comes from the relic density.

%%%%%%%%%%%%%%%%%%%
%%%%%%%%%%%%%%%%%%%

\subsection{Collider physics }
Here we briefly discuss collider search for the leptoquarks.
The leptoquarks can be produced via QCD process, $pp \to S_{LQ_{1(2)}} S_{LQ_{1(2)}}^*$, at the LHC where the production cross section is determined by their masses.
The decay of the leptoquarks is induced by the Yukawa coupling in Eq.~(\ref{eq:Yukawa}) such that 
\begin{align}
S_{LQ_1} \to  \ell_i u_j ( \nu_i d_j), \quad S_{LQ_2} \to d_i N_j. 
\end{align} 
The decay widths are given by
\begin{align}
\Gamma (S_{LQ_1} \to \ell_i u_j ( \nu_i d_j)) &= \frac{\left[ (y_L)_{ij} \right]^2 m_{S_{LQ_1}}}{8 \pi}  \left(1 - \frac{m_{\ell_i (\nu_i)}^2}{m_{LQ_1}^2} - \frac{m_{u_j(d_j) }^2}{m_{LQ_1}^2}  \right) \lambda(m_{\ell_i (\nu_i)}^2/m_{LQ_1}^2, m_{u_j(d_j)}^2/m_{LQ_1}^2)^{1/2} ,\nonumber \\
\Gamma (S_{LQ_2} \to d_i N_j ) &= \frac{\left[ (y_S)_{ij} \right]^2 m_{S_{LQ_2}}}{8 \pi}  \left(1 - \frac{m_{d_i}^2}{m_{LQ_2}^2} - \frac{m_{N_j }^2}{m_{LQ_2}^2}  \right) \lambda(m_{d_i}^2/m_{LQ_2}^2, m_{N_j}^2/m_{LQ_2}^2)^{1/2},
\end{align}
where $\lambda(x,y) = 1 +x^2+y^2 - 2x -2y -2xy$, and we take active neutrino mass as zero.
To see the tendency of branching ratio (BR), we apply the parameter sets satisfying all the phenomenological constraints which are obtained by numerical analysis in Sec.~\ref{sec:numerical}.

\begin{figure}[t]
\begin{center}
\includegraphics[width=70mm]{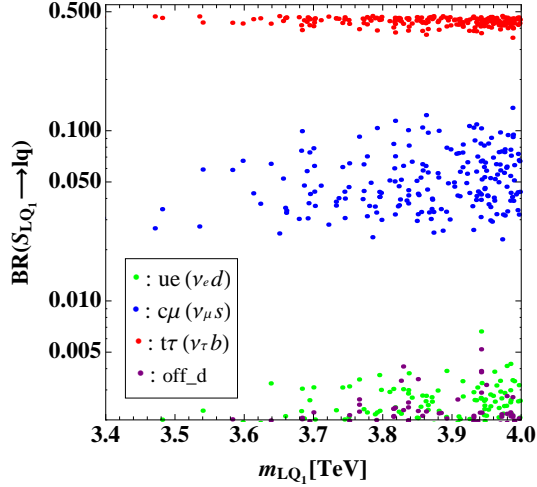}
\caption{The branching ratio of the leptoquark $S_{LQ_1}$ where the parameter sets satisfying all the constraints are applied.}
\label{fig:BR}
\end{center}
\end{figure}

In Fig.~\ref{fig:BR}, we show the BRs for $S_{LQ_1}$ and $S_{LQ_2}$ as a function of their masses.
We find that $S_{LQ_1}$ mainly decays into  $t \tau$ and $\nu b$ channels with the same BR, while $c \mu$ and $\nu s$ channels have subdominant BR.
Then the BR for the final state $\mu^+ \mu^- c \bar c$ is $\lesssim 0.3 \%$ for $S_{LQ_1}$ pair production.
Thus the $S_{LQ_1}$ in our preferred mass region is free from current experimental constraints by the channel~\cite{Aaboud:2016qeg,CMS:2016qhm} 
and much higher luminosity is required to search for $S_{LQ_1}$ in this mode.
It will be interesting to search for third generation specific signatures of $S_{LQ_1}$ pair production, $\tau^+ \tau^- t \bar t$ and $ \tau t b \nu$, which have much larger BR than $\mu^+ \mu^- c \bar c$ channels. 
On the other hands we find $S_{LQ_2}$ almost $100\%$ decays into $N_2 s$ channel.
Thus the signature of $S_{LQ_2}$ is $\ET + {\rm jets}$.
%%%
We note that the squark pair production with $\tilde q \to \tilde \chi^0 q $ decay mode provides similar signature as $S_{LQ_2}$.
Hence, we can estimate the lower limit of $S_{LQ_2}$ mass from the current data for squark search~\cite{CMS:2016mwj}.
From the limit for one squark case, we obtain the lower limit of the $S_{LQ_2}$ mass as up to  $\sim 450$ GeV, depending on mass degeneracy between $S_{LQ_2}$ and DM.
Therefore some of our preferred parameter region would already be excluded and most of the region could be tested in future LHC experiments.

%%%%%%%%%%%%%%%%%%%
%%%%%%%%%%%%%%%%%%%

\section{Conclusions}

In this paper, we have studied colored KNT model, in which scalar leptoquarks are introduced.
The active neutrino mass matrix is induced at three loop level where the leptoquarks propagate inside the loop. 
In addition, the lightest SM singlet Majorana fermion can be a dark matter candidate due to a discrete $Z_2$ symmetry imposed in the model.

We have carried out numerical analysis to search for allowed parameter range which is consistent with neutrino oscillation data and DM relic density.
Then the constraints from the flavor changing neutral current have been taken into account such as the flavor changing lepton decay $\ell_i \to \ell_j \gamma$, and $K^0-\bar K^0$ and $B_d^0- \bar B_d^0$ mixings.
We then find that 100 GeV scale DM and $S_{LQ_2}$ and TeV scale leptoquarks $S_{LQ_1}$ can be consistent with all the constraints, and all the coupling constants are in the perturbative regime.

Finally we have discussed collider physics regarding leptoquark production in the model.
The leptoquarks can be produced by QCD process and then decay into lepton and quark.
The branching ratio (BR) of $Z_2$ even (odd) leptoquark $S_{LQ_{1(2)}}$ is investigated with the parameter sets obtained from our numerical analysis. 
We have found that $S_{LQ_1}$ mainly decays into $t \tau$ and $\nu b$ channels with same BR while $c \mu$ and $\nu s$ channels have subdominant BR.
Thus BR for $S_{LQ_1} S_{LQ_1}^* \to \mu^+ \mu^- j j$ is around $\lesssim 0.3 \%$ and our preferred mass region is free from the constraints from the current experimental data.
In addition, the model could be also tested by searching for $S_{LQ_1}$ signals such as $\tau^+ \tau^- t \bar t$ and $t \tau b \nu$ which have much larger BR than $\mu^+ \mu^- j j$ channel.
On the other hand we find $S_{LQ_2}$ almost $100 \%$ decays into $N_2 s$ channel.
Thus the signature of $S_{LQ_2}$ is $\ET + {\rm jets}$ and we roughly estimate upper limit of the mass by using the current data for squark search such that up to  $\sim 450$ GeV, depending on mass degeneracy between the leptoquark and DM. Note that our preferred mass range is within the reach of current and/or near future LHC experiment.

%\section*{ Appendix}
%%%%%%%%%%%%%%%%%%%...

%\newpage
%%%%%%%%%%%%%%%%%%%%%%%%%%%%%%%%%%%
%\hspace{0.2cm} {\bf Acknowledgments}
%\section*{Acknowledgments}:
%\vspace{0.5cm}
\section*{Acknowledgments}
\vspace{0.5cm}
Authors would like thank Dr. Masaya Kohda for fruitful discussions.
H. O. is sincerely grateful for all the KIAS members, Korean cordial persons, foods, culture, weather, and all the other things.
The work of N.O is supported in part by the United States Department of Energy (DE-SC 0013680).
%%%%%%%%%%%%%%%%%%%%%%%%%%%%%%%%%%%
%%%%%%%%%%%%%%%%%%%%%%%%%%%%%%%%%%%

\end{document}